\documentclass[useAMS,usegraphicx]{mn2e}

\title[X-ray absorption variability in NGC~4395]{The effects of X-ray absorption variability in NGC~4395}
\author[E. Nardini \& G. Risaliti]{E.~Nardini$^{1}$\thanks{E-mail: enardini@cfa.harvard.edu}
and G.~Risaliti$^{1,2}$ \\
$^1$ Harvard-Smithsonian Center for Astrophysics, 60 Garden St., Cambridge, MA 02138, USA \\
$^2$ INAF - Osservatorio Astrofisico di Arcetri, L.go E. Fermi 5, 50125 Firenze, Italy}

\begin{document}

\date{Released Xxxx Xxxxx XX}

\pagerange{\pageref{firstpage}--\pageref{lastpage}} \pubyear{2011}

\maketitle

\label{firstpage}

\begin{abstract}
We present a new X-ray analysis of the dwarf Seyfert galaxy NGC~4395, based on two archival \textit{XMM-Newton} and \textit{Suzaku} 
observations. This source is well known for a series of remarkable properties: one of the smallest estimated black hole masses among 
Active Galactic Nuclei (of the order of $\sim$10$^5 M_{\sun}$), intense flux variability on very short time-scales (a few tens of seconds), 
an unusually flat X-ray continuum ($\Gamma \sim 1.4$ over the 2--10~keV energy range). NGC~4395 is also characterized by significant 
variations of the X-ray spectral shape, and here we show that such behaviour can be explained through the partial occultation by 
circumnuclear cold absorbers with column densities of $\sim$10$^{22}$--10$^{23}$~cm$^{-2}$. In this scenario, the primary X-ray 
emission is best reproduced by means of a power law with a standard $\Gamma \sim 1.8$ photon index, consistent with both 
the spectral slope observed at higher energies and the values typical of local AGN.
\end{abstract}


\section{Introduction}

The central regions of the dwarf spiral galaxy NGC~4395 show most of the common signatures of nuclear activity. The optical and 
ultraviolet (UV) spectra reveal prominent high-ionization forbidden lines on top of a nearly featureless continuum, and broad wings 
corresponding to gas velocities in excess of $\sim$10$^3$~km~s$^{-1}$ are detected in the permitted lines (Filippenko \& Sargent 
1989). Contrary to the objects of the same kind, the emission-line properties, the optical to X-ray variability pattern and the inferred 
accretion rate of NGC~4395 are those typical of the Seyfert class, of which this source is usually considered to represent the least 
luminous member. Different methods have been employed in the last years to derive the mass of its central black hole: the estimate 
obtained through reverberation mapping is $M_\rmn{BH} \simeq 3.6 \times 10^5 M_{\sun}$ (Peterson et al. 2005), but the lack in this 
galaxy of a significant bulge and the stringent upper limit of 30~km~s$^{-1}$ on its velocity dispersion suggest an even lower value, 
of the order of $\sim$10$^4$--10$^5 M_{\sun}$ (Filippenko \& Ho 2003). Anyhow, the engine of NGC~4395 falls somewhere 
between the stellar-mass black holes found in Galactic X-ray binaries and the supermassive black holes residing inside active 
galactic nuclei (AGN). As such, it can provide critical information about the relationship between these two populations and the 
physics of accretion systems in general. \\
In the light of all these pieces of observational evidence, NGC~4395 is a true scaled-down version of an ordinary Seyfert galaxy, 
the only difference with respect to its high-luminosity counterparts being the much smaller mass of the central black hole. On the 
other hand, the X-ray observations of this source indicate an unusual spectral hardness at $\sim$2--10~keV, and suggest a wide 
range of variations for the intrinsic photon index. The most extreme states ($\Gamma \sim 0.6$; Moran et al. 2005) are even difficult 
to interpret within the standard two-phase model (Haardt \& Maraschi 1991), posing serious questions on the production mechanism 
of the X-ray emission itself. The presence of undetected absorption effects has been frequently invoked as a possible explanation. 
Indeed, the stronger flux variability characterizing systematically the soft X-ray bands can be attributed to a complex, multi-zone 
warm absorber, whose properties have been discussed in detail in several works (Iwasawa et al. 2000; Shih, Iwasawa \& 
Fabian 2003; Dewangan et al. 2008). Here we review the two highest-quality observations of NGC~4395, performing in both 
cases an accurate time-resolved analysis mainly focused on neutral absorption, in order to test whether changes of its column 
density and/or covering factor play a role in the apparent X-ray spectral hardness of this source. 

\section{Data reduction and analysis}

The longest \textit{XMM-Newton} monitoring of NGC~4395 started on 2003 November 30, for a total duration of $\sim$113~ks. 
After the subtraction of high-background periods, the useful exposure declines to 91.4~ks. We also take into account the deep 
\textit{Suzaku} observation, which was carried out on 2007 June 2--5 over a span of $\sim$230~ks, corresponding to a net 
integration time of 101.3~ks. We have followed the standard procedures for the reduction of the event files, and extracted the 
source and background spectra from circular regions with radii of 30\arcsec (\textit{XMM-Newton}) and 2\arcmin (\textit{Suzaku}). 
In the first case, for the sake of clarity only EPIC-pn data are presented here and plotted in the figures, even though the MOS 
spectra have been checked throughout and give fully consistent results; in the second one, instead, only the data from the 
front-illuminated detectors of the X-ray imaging spectrometer (XIS) have been examined, after merging the XIS0 and XIS3 
spectra. The spectral analysis has been performed using the \textsc{xspec} v12.6 fitting package. All the uncertainties are 
given at the 90 per cent confidence level ($\Delta \chi^2 = 2.71$) for the single parameter of interest. 

\subsection{The \textit{XMM-Newton} observation}
\begin{figure}
\includegraphics[width=8.5cm]{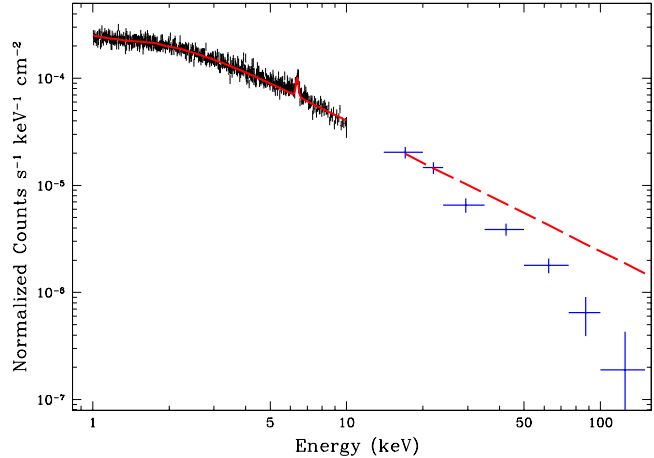}
\caption{X-ray emission of NGC~4395 above 1~keV and best fit (red solid line) of the entire \textit{XMM-Newton} observation. The 
extrapolation of this benchmark model at higher energies (red dashed line) is compared with the average spectrum extracted from the 
\textit{Swift}/BAT 58-month catalog, whose shape is perfectly described by a simple power law with $\Gamma \simeq 2.0 (\pm 0.2)$.} 
\label{bm}
\end{figure}

\begin{figure}
\includegraphics[width=8.5cm]{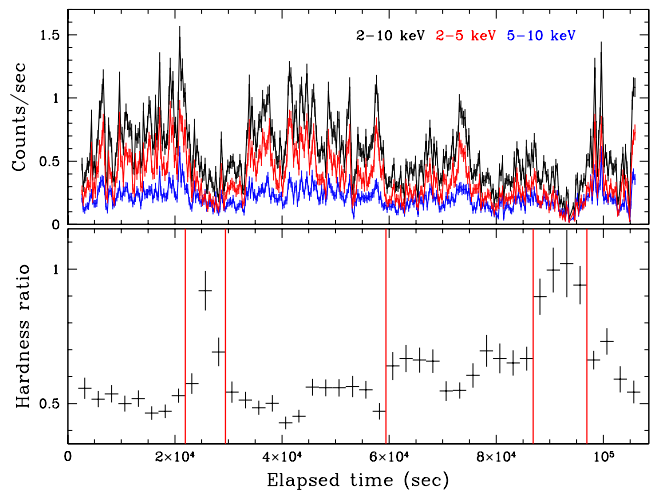}
\caption{Top panel: 2--10~keV light curve of NGC~4395 during the \textit{XMM-Newton} observation (in black). The behaviour of the source at 
2--5 (red) and 5--10~keV (blue) is also shown separately, adopting a bin width of 250~s. Bottom panel: 5--10 over 2--5~keV hardness ratio, 
with the time resolution degraded by a factor of ten. The vertical red lines define the six intervals over which individual spectra have been extracted 
in our time-resolved analysis.}
\label{lc}
\end{figure}

\begin{figure}
\includegraphics[width=8.5cm]{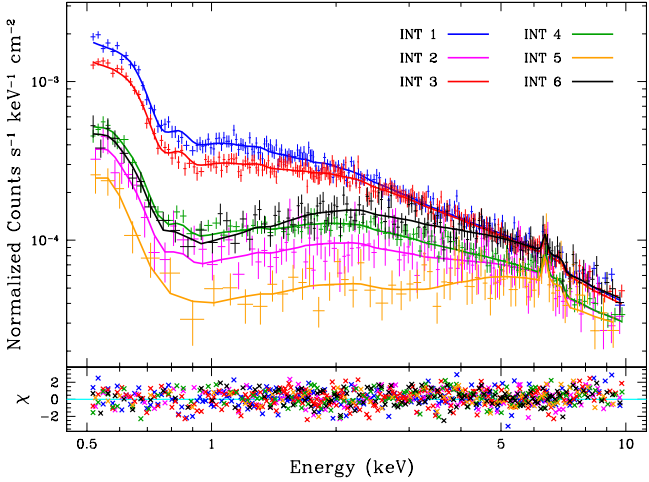}
\caption{Top panel: spectra and best-fitting models for the six \textit{XMM-Newton} intervals. Bottom panel: residuals in units of $\sigma$. 
The data are rebinned for plotting purposes only.}
\label{xmm}
\end{figure}

We first restricted our analysis to the energies above 1~keV, following the customary approach of fitting the spectrum averaged 
over the whole observation to obtain a benchmark model. We achieved a fully acceptable fit ($\chi_\nu^2 \simeq 1.03$ for 1390 
degrees of freedom, with no obvious structure in the residuals) through a simple model consisting of a power law with photon index 
$\Gamma \simeq 1.18$, a narrow iron emission line with equivalent width $\rmn{EW} = 91(\pm 30)$~eV, and a partial neutral 
absorber whose column density and covering fraction are $N_\rmn{H} \simeq 1.3 \times 10^{22}$~cm$^{-2}$ and 
$f_\rmn{cov} \simeq 0.6$, respectively.\footnote{Galactic absorption has been frozen throughout at 
$N_\rmn{H} = 1.8 \times 10^{20}$~cm$^{-2}$ (Kalberla et al. 2005).} No reflection component is strictly required for the continuum: 
however, by adding a \textsc{pexrav} model (Magdziarz \& Zdziarski~1995) for physical consistency and forcing its strength to 
match the width of the narrow iron line,\footnote{Assuming solar iron abundance, this implies that continuum reflection and line 
emission are associated with the same reprocessing component (e.g. George \& Fabian 1991).} the fit is slightly improved 
($\Delta \chi^2 \simeq -8$ with the loss of one d.o.f.) and the power-law photon index steepens to $\simeq 1.24$. Both cases 
above yield a formally satisfactory fit, yet the flat slope of the intrinsic continuum deserves a thorough investigation. This result 
would actually confirm one of the most remarkable features of NGC~4395, which had been previously caught by \textit{Chandra} 
even in harder states (see Moran et al. 2005). In any case, the typical value of $\Gamma \sim 1.3$--1.5 is still unusually low, 
compared with both the distribution of photon indices found in local AGN (e.g. Bianchi et al. 2009) and the average 15--150~keV 
spectrum of the source measured by the Burst Alert Telescope (BAT) onboard \textit{Swift} ($\Gamma \simeq 2.0$; Fig.~\ref{bm}). \\
A viable explanation for such a discrepancy is the presence of a complex (variable) absorber, strongly modifying the observed 
spectral shape of NGC~4395 below $\sim$10~keV. The extrapolation of the basic model down to 0.5 keV supports this working 
assumption: the apparent extra emission at 0.5--0.7~keV, in fact, hints at absorption effects in the $\sim$0.7--1.5~keV range rather 
than at a genuine soft excess like the one detected in a large fraction of Seyfert galaxies (e.g. Boller, Brandt \& Fink 1996; Porquet 
et al. 2004). Moreover, a clearly time-dependent behaviour of the X-ray source is revealed not only by the overall flux light curve but 
also by the hardness ratio (HR) evolution. It is well established that NGC~4395 is characterized by strong variations of the X-ray 
flux on time-scales as short as a few hundreds of seconds (Fig.~\ref{lc}, upper panel): the fractional rms variability amplitude during 
the \textit{XMM-Newton} observation under review is exceptionally high ($\sim$90 per cent at 0.5--2~keV; see Vaughan et al. 2005). 
On top of these fluctuations of the intrinsic X-ray brightness, mainly related to the workings of the primary source, significant 
\textit{spectral} variations are also evident from the visual inspection of the HR light curve (Fig.~\ref{lc}, lower panel). The observed 
HR pattern can be accounted for through either changes of the column density or, alternatively, oscillations of the continuum slope: 
this proves that a time-averaged spectral analysis is not sufficient to fully understand the nature of the X-ray emission of NGC~4395. 
The hardness ratio is defined here as the ratio between the 5--10~keV and the 2--5~keV flux, thus it is especially sensitive to values 
of $N_\rmn{H}$ of the order of 10$^{22}$--10$^{23}$~cm$^{-2}$. As already mentioned, the spectral complexity and the larger fractional 
rms variability below $\sim$2~keV are likely due instead to a multi-zone warm absorber, whose presence has been revealed since the 
early \textit{ASCA} observations of the source (Iwasawa et al. 2000; Shih et al. 2003). We anyway note that this is not a critical point, 
since the HR evolution is only used to select the appropriate intervals for the subsequent time-resolved spectral analysis, which is crucial 
to validate or dismiss the suggested interpretation. \\
The shape of the HR light curve reveals two periods of sudden spectral hardening around $\sim$25 and 90~ks; this hints at increased 
opacity, which would mainly affect the 2--5~keV band. According to Fig.~\ref{lc}, six different regimes of the hardness ratio can be roughly 
defined. In this framework, the minor HR fluctuations on very short time-scales can be due to a flickering of the photon index in response 
to variations of the physical conditions in the disc/corona system, but the statistics is not enough for a complete investigation of these 
aspects. Regardless of this, our aim is to check whether a reasonable configuration of the neutral absorber allows us to recover a steeper 
intrinsic photon index for the X-ray emission of NGC~4395. \\
In order to obtain an adequate description of the entire 0.5--10~keV spectral range, we included in the reference model above also an 
\textsc{apec} component for the soft thermal emission (Smith et al. 2001) and a two-zone warm absorber, whose complex effects have 
been already pointed out in several previous works (see also Dewangan et al. 2008). In our analysis we have assumed that the soft 
emission, warm absorption and reflection features do not vary in the course of a $\sim$100~ks long observation: since the light crossing 
time over a distance of a gravitational radius ($r_\rmn{g} = GM_\rmn{BH}/c^2$) is of the order of one second, the reflection component is 
expected to change significantly in response to the primary continuum only if the scattering material is located well within $\sim$10$^5 r_\rmn{g}$ 
from the centre (as in the disc reflection scenarios; e.g. Nardini et al. 2011). Similarly, the soft emission likely arises from a very extended region, 
while the warm absorber, in principle, could lie much closer to the X-ray source (but see Blustin et al. 2005): if this is the case, we are just 
sampling its average properties.\footnote{The warm absorbers have been modelled with \textsc{absori}, adopting the nominal \textsc{xspec} 
values (with solar abundances) and a full covering. We are not interested in a more detailed study of this component here, hence its nature is 
not explored further.} On the other hand, $\Gamma$, $N_\rmn{H}$ and $f_\rmn{cov}$ were initially left free to evolve among the different intervals. 
\begin{figure}
\includegraphics[width=8.5cm]{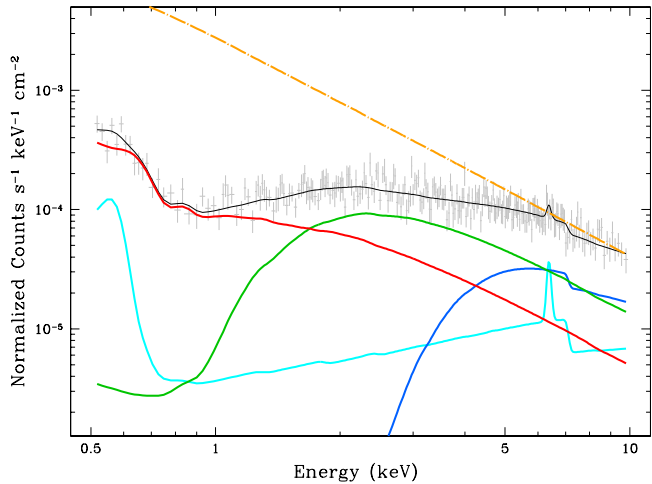}
\caption{Illustration of the different model components for the \textit{XMM-Newton} spectrum extracted from interval 6 (shaded in the background along 
with its best fit). The soft emission plus reflection features (cyan curve) are assumed to be constant over the entire observation; the spectral variations 
are therefore due to the relative weight of the three fractions of the intrinsic power law (dot-dashed orange line), transmitted through the warm absorber 
only and an additional neutral column $N_\rmn{H}$(1) or $N_\rmn{H}$(2), and plotted respectively in red, green and blue.}
\label{sc}
\end{figure}
In spite of the possible degeneracies, it turns out that the photon index and the column density are subject to very limited variations: in particular, 
$\Delta \Gamma < 0.2$. This suggests that $\Gamma$ and $N_\rmn{H}$ can be roughly treated as constant parameters as well, hence their 
values have been tied in all the six spectra: only the covering fraction of the cold absorber is allowed to vary with time. \\ 
On sheer statistical grounds, this model already gives a very good fit ($\chi^2_\nu/\rmn{d.o.f.}=0.994/1788$, with $\Gamma \simeq 1.90$ 
and $N_\rmn{H} \sim 2.1 \times 10^{22}$~cm$^{-2}$), proving that, in first approximation, $f_\rmn{cov}$ alone would be able to account for 
all the observed spectral variability: its qualitative trend across the six intervals, in fact, fairly correlates with the shape of the HR light 
curve. There are some problems with the reflection efficiency, though. It is not obvious how to assess the latter quantity in the time-resolved 
analysis, anyway by comparing the \textsc{pexrav} amplitude to the highest flux level of the power-law continuum we obtain a reflection 
strength of $R \sim 7.5$, which is completely inconsistent with the moderate equivalent width of the companion iron line (George \& Fabian 
1991). Moreover, when extrapolated at higher energies, such component would give rise to a very flat spectrum with a prominent Compton 
hump at $\sim$30~keV, which is not detected by \textit{Swift}/BAT. Given that the time-scales are extremely different, it is still possible that 
the source is caught in a state of large reflection during the \textit{XMM-Newton} observation, but in this case the fluorescent iron line should 
be much stronger as well. Forcing $R$ to have a standard value delivers a poor spectral description, and no improvement is achieved 
through the self-consistent \textsc{reflionx} table models (Ross \& Fabian 2005). \\
As a consequence, we attempted to introduce a second partial covering cold absorber, assuming again that $\Gamma$ and $N_\rmn{H}$(1,2) 
are constant, and $f_\rmn{cov}$(1,2) variable. The best fit quality is significantly improved ($\chi^2_\nu/\rmn{d.o.f.}=0.961/1781$), while the 
photon index $\Gamma \simeq 1.86$ is still in excellent agreement with the average high-energy slope of the source. The presence of an 
additional absorber with column density of $\sim$10$^{23}$~cm$^{-2}$ removes all the previous limitations, reducing substantially the required 
strength of the reflection continuum (now $R \sim 1.2$). This is also clear from Fig.~\ref{sc}, where the different model components are disentangled 
to display their relative contribution in a single time segment. The basic parameters of this final model are listed in Table~\ref{t1}, while the six 
individual spectra are shown in Fig.~\ref{xmm}. 
\begin{table}
\caption{Main parameters of the \textit{XMM-Newton} and \textit{Suzaku} best-fitting models, common to all the intervals on which 
our time-resolved spectral analysis has been performed. $\Gamma$: photon index; $kT$: temperature of the soft emission in eV; 
$N_W$, $\xi_W$: column density in cm$^{-2}$ and ionization parameter in erg~cm~s$^{-1}$ of the warm absorption components; 
EW$_\rmn{Fe}$: time-averaged equivalent width of the 6.4-keV iron line in eV; $R$: strength of the reflected continuum; $N_\rmn{H}$: 
column density of the neutral absorbers in cm$^{-2}$; $F_\rmn{obs}$, $F_\rmn{int}$: average observed and intrinsic 0.5--10~keV 
flux in erg~cm$^{-2}$~s$^{-1}$.}
\label{t1}
\begin{tabular}{l@{\hspace{40pt}}c@{\hspace{40pt}}c}
\hline
Obs. & \textit{XMM-Newton} & \textit{Suzaku} \\
$\Gamma$ & $1.86 \pm 0.06$ & 1.74$^{+0.12}_{-0.14}$ \\
$kT$ & 41$^{+6}_{-11}$ & 41$^*$ \\
$N_W$(1) & 0.67$^{+0.03}_{-0.04} \times$10$^{22}$ & 0.51$^{+0.18}_{-0.17} \times$10$^{22}$ \\
$\xi_W$(1) & 3.5$^{+0.9}_{-0.4}$ & 5.9$^{+8.8}_{-3.0}$ \\
$N_W$(2) & 1.90$^{+0.45}_{-0.35} \times$10$^{22}$ & 1.90$^* \times$10$^{22}$ \\
$\xi_W$(2) & 368$^{+113}_{-83}$ & 368$^*$ \\
EW$_\rmn{Fe}$ & 64$^{+35}_{-32}$ & $81 \pm 53$ \\
$R$ & $1.2 \pm 1.1$ & 1.4$^{+2.9}_{-1.3}$ \\
$N_\rmn{H}$(1) & 1.68$^{+0.23}_{-0.28} \times$10$^{22}$ & 1.07$^{+0.40}_{-0.32} \times$10$^{22}$ \\
$N_\rmn{H}$(2) & 2.55$^{+0.51}_{-0.46} \times$10$^{23}$ & 0.92$^{+0.85}_{-0.42} \times$10$^{23}$ \\
$F_\rmn{obs}$ & $6.25 \times 10^{-12}$ & $5.54 \times 10^{-12}$ \\
$F_\rmn{int}$ & $1.36 \times 10^{-11}$ & $9.15 \times 10^{-12}$ \\
$\chi^2_\nu$ & 1711/1781 & 1095/1146 \\
\hline
\end{tabular}
$^*$: frozen value.
\end{table}

\subsection{The \textit{Suzaku} observation}

\begin{figure}
\includegraphics[width=8.5cm]{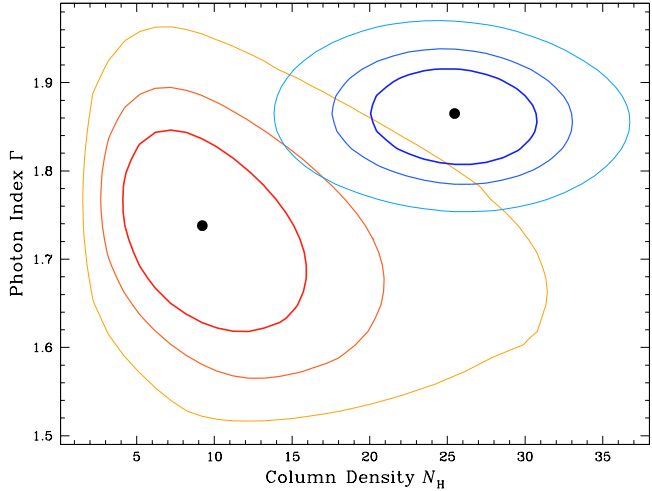}
\caption{Confidence contours at the 68, 90 and 99 per cent level in the $\Gamma$--$N_\rmn{H}$(2) space for both the \textit{XMM-Newton} and the 
\textit{Suzaku} observation. The photon index is well constrained and the two measures are in good agreement. ($N_\rmn{H}$ is in units of 
10$^{22}$~cm$^{-2}$).}
\label{gnh}
\end{figure}

\begin{figure}
\includegraphics[width=8.5cm]{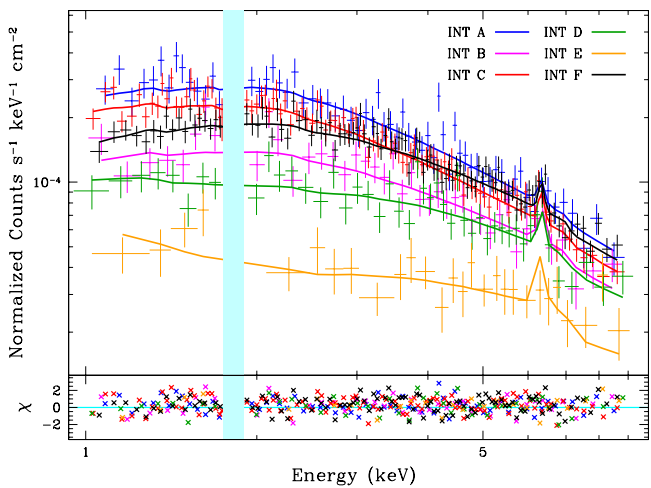}
\caption{Same as Fig.~\ref{xmm} for the \textit{Suzaku} observation. Only the 1--9~kev range is plotted for clarity, while the 1.75--1.9~keV 
region is excluded from the fits due to calibration uncertainties around the instrumental silicon absorption edge.}
\label{szk}
\end{figure}

In spite of the promising indications, the cold absorption scenario needs to be tested on different X-ray observations of NGC~4395 to be 
definitely regarded as reliable for this source. Indeed, a complex configuration consisting of both a partially and a fully covering neutral absorber 
(plus three warm components) has been shown to provide an adequate fit of the time-averaged spectrum reviewed here, and has been 
also applied with success to the short \textit{XMM-Newton} snapshots of NGC~4395 (Dewangan et al. 2008). In general, however, the 
number of counts collected during the archival observations is not sufficient to derive solid and independent constraints on our model. 
We have therefore analysed the only other long, high-quality monitoring of NGC~4395 that is available to date, obtained by \textit{Suzaku}  
and fully discussed by Iwasawa, Tanaka \& Gallo (2010). In the latter work,  the time-averaged spectrum is described by means of a 
power-law continuum with $\Gamma \sim 1.4$, modified by a warm absorber with ionization parameter $\xi_{W} \sim 35$ erg~cm~s$^{-1}$ 
and column density $N_{W} \simeq 2 \times 10^{22}$~cm$^{-2}$; a narrow iron line and neutral absorption in moderate excess of the Galactic 
amount are also involved. By dividing the whole observation into six segments of equal length, and allowing $\Gamma$, $\xi_W$ and 
$N_W$ to vary in turn among the six spectra, the authors suggest that a change of the photon index might account for most of the spectral 
variability of the X-ray source. None the less, the 15--35~keV \textit{Suzaku}/PIN spectrum reveals a slope of $\sim 2.2(\pm 0.5)$, consistent 
with the \textit{Swift}/BAT measure and possibly indicative of some undetected absorption component at lower energies. \\
On the wake of this conjecture, we have also performed a time-resolved analysis, yet choosing the time intervals on the grounds of the HR light 
curve, as usual. The 5--10 over 2--5 keV hardness ratio follows a trend which is qualitatively very similar to that shown in Fig.~\ref{lc} (see e.g. 
the X-ray colour HR3 in Iwasawa et al. 2010): again, six different periods have been identified. We have adopted the same model defined above. 
However, the soft thermal emission and the high-ionization component of the warm absorber cannot be firmly constrained, hence their properties 
have been frozen to the values obtained in the \textit{XMM-Newton} study. In spite of this assumption we are able to recover an excellent fit to 
the data, with $\chi^2_\nu/\rmn{d.o.f.}=0.956/1146$. The photon index of the primary continuum is slightly flatter ($\Gamma \sim 1.75$), but its 
uncertainty combined with that on $N_\rmn{H}$(2) is rather large due to the lower data quality (see Fig.~\ref{gnh}). Indeed, if $\Gamma$ is bound 
to the \textit{XMM-Newton} best-fitting measure the variation in terms of statistical goodness is very limited, being $\Delta \chi^2 \simeq +3$. 
According to an $F$-test, the probability of chance improvement when $\Gamma$ is included among the free parameters amounts to $\sim$8.3 
per cent. In any case, the estimated photon index is again well above the value obtained by describing the time-averaged spectrum with a 
simpler model. Our results are summarized in Tables~\ref{t1} and \ref{t2}, and the six \textit{Suzaku} spectra are shown in Fig.~\ref{szk}. 
\begin{table}
\caption{Time evolution of the covering fractions during the two observations and $\chi^2_\nu$ decomposition 
over the single intervals. 
By definition, $f_\rmn{cov}$(0) is the fraction of the X-ray source subject to warm absorption only.}
\label{t2}
\begin{tabular}{ccccc}
\hline
INT & $f_\rmn{cov}$(0) & $f_\rmn{cov}$(1) & $f_\rmn{cov}$(2) & $\chi^2_\nu$ \\ 
1 & 0.62$\pm$0.11 & $< 0.03$ & 0.38$\pm$0.11 & 440/424 \\ 
2 & 0.14$\pm$0.05 & 0.26$\pm$0.09 & 0.60$\pm$0.12 & 118/106 \\ 
3 & 0.49$\pm$0.10 & 0.20$\pm$0.06 & 0.31$\pm$0.12 & 488/473 \\ 
4 & 0.22$\pm$0.06 & 0.31$\pm$0.09 & 0.47$\pm$0.13 & 395/428 \\ 
5 & 0.07$\pm$0.04 & 0.14$\pm$0.07 & 0.79$\pm$0.09 & 55/55 \\ 
6 & 0.13$\pm$0.03 & 0.34$\pm$0.08 & 0.53$\pm$0.10 & 215/235 \\ 
\hline
A & 0.36$\pm$0.19 & 0.64$\pm$0.20 & $< 0.16$ & 117/128 \\ 
B & 0.28$\pm$0.18 & 0.48$\pm$0.23 & 0.24$\pm$0.20 & 92/81 \\ 
C & 0.40$\pm$0.17 & 0.60$\pm$0.18 & $< 0.11$ & 375/393 \\ 
D & 0.26$\pm$0.17 & 0.24$\pm$0.20 & 0.50$\pm$0.22 & 76/91 \\ 
E & 0.38$\pm$0.23 & $< 0.25$ & 0.62$\pm$0.25 & 40/27 \\ 
F & 0.22$\pm$0.12 & 0.50$\pm$0.15 & 0.28$\pm$0.13 & 405/388 \\ 
\hline
\end{tabular}
\end{table}

\section{Discussion}

\begin{figure}
\includegraphics[width=8.5cm]{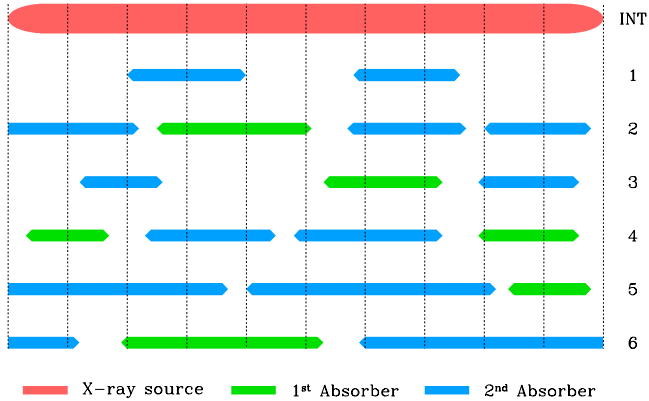}
\caption{Sketch of the simplest structure of the X-ray cold absorber, consisting of individual clouds at the same distance from the X-ray source but 
with different column densities. The covering factors across the six \textit{XMM-Newton} intervals have been chosen to reproduce the values listed 
in Table~\ref{t2}. The number of clouds simultaneously crossing the line of sight and their size are limited by the rapid changes of $f_\rmn{cov}$. 
Even a single blob with irregular shape and non-uniform physical properties is consistent with the observed variability pattern. (The observer is 
located towards the bottom of the page).}
\label{blr}
\end{figure}

\begin{figure}
\includegraphics[width=8.5cm]{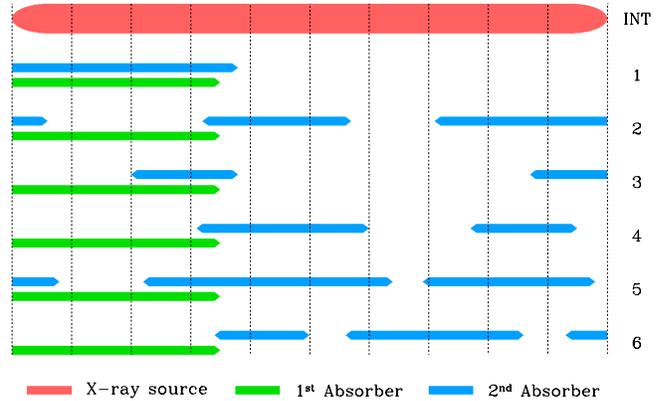}
\caption{Same as Fig.~\ref{blr}, but an alternative configuration is assumed: the absorption component with lower column density has a constant 
covering fraction of $0.35$ and is external to the system of clouds responsible of the X-ray spectral variability (see the discussion in the text).}
\label{abs}
\end{figure}

The cold absorption model provides a good interpretation of the spectral variability of the X-ray source during both the \textit{XMM-Newton} 
and \textit{Suzaku} observations. Moreover, considering the entries of Table~\ref{t1}, the basic physical quantities appear to be in fair agreement. 
This is a strong confirmation of the validity of the scenario explored in this work, where the changes are driven by the evolution of the covering 
fractions. The $f_\rmn{cov}$(1) and $f_\rmn{cov}$(2) progression is listed in Table~\ref{t2}. Due to the smaller uncertainties, we focus our 
discussion on the \textit{XMM-Newton} case. We first point out that our model is defined in such a way that mutually exclusive regions of the 
X-ray source are affected by the two cold absorbers: in other words, we only consider a single-layer configuration of the neutral gas along the 
line of sight. All the $f_\rmn{cov}$ sequences should be taken with some caution, though. The low ionization stage in one of the warm absorption 
component impairs to some extent the determination of $f_\rmn{cov}$(1), while possible changes in the reflection strength on that of $f_\rmn{cov}$(2). 
However, even from the face values obtained under our assumptions, some interesting considerations can be drawn. It is now 
$f_\rmn{cov}(2, t) = \lbrace 0.38, 0.60, 0.31, 0.47, 0.79, 0.53 \rbrace$ that shows a tighter correlation with the HR light curve: the peaks during 
periods 2 and 5 might be explained in terms of \textit{eclipses} from individual \textit{clouds} (e.g. Lamer, Uttley \& McHardy 2003; Risaliti et al. 
2007). Given that the duration of such events is $\sim$10~ks, a typical dimension of the X-ray source of $\sim$10$^2 r_\rmn{g}$ corresponds 
to a transverse velocity of the putative clouds of $\sim$10$^3$~km s$^{-1}$, placing the obscuring gas at the broad-line region scale. The size 
and the shape of these blobs are not precisely known. Yet, our time-resolved analysis suggests that different column densities are present along 
the line of sight at the same time. The range of variations in $f_\rmn{cov}$ implies that the number of intervening clouds is very limited (a few at 
most, see Fig~\ref{blr}), and that their size is comparable to the dimensions of the X-ray source. The extreme case entails a single irregular and 
inhomogeneous absorber. \\
The physical situation is then expected to be rather complex. In this view, the X-ray absorbers that we have included should be only regarded 
as a linear combination of the \textit{real} ones, whose exact geometrical structure cannot be probed with the present data quality. The large 
disparity (roughly an order of magnitude) between the column densities involved, and specifically the fact that 
$N_\rmn{H}(1) < \sigma_{N_\rmn{H}}(2)$, hints at a multi-layer configuration of the cold absorber, where the two phases have a different location 
and are superimposed in part on one another when seen in projection. Such a scheme can be envisaged as in Fig.~\ref{abs}: only the broad-line 
region component, with $N_\rmn{H} \sim 10^{23}$~cm$^{-2}$, has a time-dependent covering fraction ($f_\rmn{cov} \sim 0.3$--0.8). Conversely, 
the absorption system with lower column density is constant (or at least variable over much longer scales), and can be associated with a more 
distant gaseous component, within a narrow-line or torus-like region. Indeed, this would be consistent with both the optical classification of 
NGC~4395 as a type 1.5--1.8 Seyfert galaxy (Ho et al. 1997; Panessa et al. 2006) and the large covering factor of the emission-line regions 
with respect to the central source (Kraemer et al. 1999). The broad- and narrow-line regions are expected to be responsible for some of the UV 
to X-ray absorption detected in NGC~4395 (Crenshaw et al. 2004), and evidence for the identification of the rapidly variable X-ray absorber 
with the broad-line emitting clouds have been recently found in other Seyfert galaxies (e.g. Maiolino et al. 2010; Risaliti et al. 2011). In this 
perspective, the frequency and amplitude of the variations in $N_\rmn{H}$ and/or $f_\rmn{cov}$ would be linked to the degree of clumpiness 
of the circumnuclear environment at the different scales. \\
Once the \textit{Suzaku} data are taken into account, all the considerations made above still hold, even though the uncertainties on the key 
parameters are quite large, and the smaller difference between $N_\rmn{H}$(1) and $N_\rmn{H}$(2) makes the covering fraction patterns less 
meaningful. Apart from the speculations on the physical and geometrical structure of the cold absorption system, the scenario outlined in this 
work has the other great advantage of strengthening the correlation between NGC~4395 and the high-luminosity AGN population. First, it gives 
reasons for the X-ray spectral hardness of this source, reconciling the estimate of its photon index with the typical values found among 
AGN; secondly, neutral absorption variability systematically occurs in a significant fraction of active galaxies, the prototypical case being another type 
1.8 Seyfert, NGC~1365 (e.g. Risaliti et al. 2009). Our findings then represent a further point of contact between NGC~4395 and its more massive 
and luminous counterparts. Incidentally, it is also worth noting that the intrinsic 0.5--10~keV emission of NGC~4395 implied in this cold absorption 
scheme is larger than the observed one by just a factor of $\sim$2 (Table~\ref{t1}). Taking advantage of the \textit{Swift}/BAT spectral constraints, 
the resulting 0.5--100~keV luminosity is $\sim$6--$8 \times 10^{40}$~erg s$^{-1}$, which exceeds the usual estimates of the bolometric luminosity (see 
also Moran et al. 2005; Iwasawa et al. 2010). Depending of the exact value of the black hole mass, the Eddington ratio of NGC~4395 could be 
much closer to $\sim$0.01 than previously thought. 

%

\section{Conclusions}

We have discussed a possible interpretation of the X-ray spectral hardness usually observed in the low-luminosity active galaxy 
NGC~4395. This source harbours in its centre a black hole with estimated mass of $\sim$10$^5 M_{\sun}$, and it is one of the few 
known objects whose study can shed light on the links between the physics of accretion processes in Galactic black hole binaries 
and AGN. In spite of being a genuine Seyfert galaxy in many respects, NGC~4395 remains a somewhat puzzling source because 
of the inferred flatness of its primary X-ray continuum. The existence of complex absorption effects has often been proposed as a likely 
explanation. Here we have provided for the first time an example of these effects based on observational evidence, by reviewing 
the two highest-quality looks of NGC~4395 taken by \textit{XMM-Newton} and \textit{Suzaku}: in both cases, a time-resolved analysis 
shows that the spectral evolution of the source can be interpreted by means of a two-phase neutral absorber with variable covering 
factor. As a first approximation, this cold absorber is identified with the system of broad-line clouds, allowing for a double-peaked 
$N_\rmn{H}$ distribution. The low column density component can otherwise be attributed to an external narrow-line or torus-like 
region, with nearly constant $f_\rmn{cov}$. This is presumably an oversimplification of the physical and geometrical structure of the 
circumnuclear environment (which is known to comprise also a complex, multi-zone warm absorber), but even the existence of a 
single partial-covering cold screen cannot be completely ruled out on statistical grounds. \\
We stress that the absorption variability scenario presented here is not unique, and different models based on intrinsically flat X-ray 
continua (with significant changes of either the photon index or the reflection strength) can describe the observed behaviour of this 
source equally well. This interpretation, however, fits into the analogy between the properties of NGC~4395 and those of \textit{standard} 
high-luminosity Seyfert galaxies, many of which are systematically affected by X-ray absorption variability due to the clumpiness of 
their circumnuclear regions. Moreover, it allows us to retrieve in a natural way a $\Gamma \simeq 1.8$ photon index below 10~keV, in 
perfect agreement with the \textit{Swift}/BAT spectral slope and the usual values measured among AGN. No alternative explanation 
would then be required for the intrinsic 2--10~keV flatness of NGC~4395. 

\section*{Acknowledgments}

This work has been partly supported by NASA grant NNX08AN48G. We thank the anonymous referee for constructive and 
useful comments which significantly improved the content of this paper.



\label{lastpage}


\begin{thebibliography}{}
\bibitem[\protect\citeauthoryear{Bianchi et 
al.}{2009}]{2009A&A...495..421B} Bianchi S., Guainazzi M., Matt G., Fonseca Bonilla N., Ponti G., 2009, A\&A, 495, 421
\bibitem[\protect\citeauthoryear{Blustin et 
al.}{2005}]{2005A&A...431..111B} Blustin A.~J., Page M.~J., Fuerst S.~V., Branduardi-Raymont G., Ashton C.~E., 2005, A\&A, 431, 111
\bibitem[\protect\citeauthoryear{Boller, Brandt, 
\& Fink}{1996}]{1996A&A...305...53B} Boller T., Brandt W.~N., Fink H., 1996, A\&A, 305, 53
\bibitem[\protect\citeauthoryear{Crenshaw et 
al.}{2004}]{2004ApJ...612..152C} Crenshaw D.~M., Kraemer S.~B., Gabel 
J.~R., Schmitt H.~R., Filippenko A.~V., Ho L.~C., Shields J.~C., Turner 
T.~J., 2004, ApJ, 612, 152
\bibitem[\protect\citeauthoryear{Dewangan et 
al.}{2008}]{2008ApJ...689..762D} Dewangan G.~C., Mathur S., Griffiths 
R.~E., Rao A.~R., 2008, ApJ, 689, 762
\bibitem[\protect\citeauthoryear{Filippenko 
\& Sargent}{1989}]{1989ApJ...342L..11F} Filippenko A.~V., Sargent W.~L.~W., 1989, ApJ, 342, L11
\bibitem[\protect\citeauthoryear{Filippenko 
\& Ho}{2003}]{2003ApJ...588L..13F} Filippenko A.~V., Ho L.~C., 2003, ApJ, 588, L13
\bibitem[\protect\citeauthoryear{George 
\& Fabian}{1991}]{1991MNRAS.249..352G} George I.~M., Fabian A.~C., 1991, MNRAS, 249, 352
\bibitem[\protect\citeauthoryear{Haardt 
\& Maraschi}{1991}]{1991ApJ...380L..51H} Haardt F., Maraschi L., 1991, ApJ, 380, L51
\bibitem[\protect\citeauthoryear{Ho et al.}{1997}]{1997ApJS..112..391H} Ho 
L.~C., Filippenko A.~V., Sargent W.~L.~W., Peng C.~Y., 1997, ApJS, 112, 391
\bibitem[\protect\citeauthoryear{Iwasawa, Tanaka, 
\& Gallo}{2010}]{2010A&A...514A..58I} Iwasawa K., Tanaka Y., Gallo L.~C., 2010, A\&A, 514, A58 
\bibitem[\protect\citeauthoryear{Kalberla et 
al.}{2005}]{2005A&A...440..775K} Kalberla P.~M.~W., Burton W.~B., Hartmann D., Arnal E.~M., Bajaja E., Morras R., P{\"o}ppel W.~G.~L., 2005, A\&A, 440, 775
\bibitem[\protect\citeauthoryear{Kraemer et 
al.}{1999}]{1999ApJ...520..564K} Kraemer S.~B., Ho L.~C., Crenshaw D.~M., 
Shields J.~C., Filippenko A.~V., 1999, ApJ, 520, 564
\bibitem[\protect\citeauthoryear{Lamer, Uttley, 
\& McHardy}{2003}]{2003MNRAS.342L..41L} Lamer G., Uttley P., McHardy I.~M., 2003, MNRAS, 342, L41
\bibitem[\protect\citeauthoryear{Magdziarz 
\& Zdziarski}{1995}]{1995MNRAS.273..837M} Magdziarz P., Zdziarski A.~A., 1995, MNRAS, 273, 837
\bibitem[\protect\citeauthoryear{Maiolino et 
al.}{2010}]{2010A&A...517A..47M} Maiolino R., et al., 2010, A\&A, 517, A47
\bibitem[\protect\citeauthoryear{Moran et al.}{2005}]{2005AJ....129.2108M} 
Moran E.~C., Eracleous M., Leighly K.~M., Chartas G., Filippenko A.~V., Ho 
L.~C., Blanco P.~R., 2005, AJ, 129, 2108
\bibitem[\protect\citeauthoryear{Nardini et 
al.}{2011}]{2011MNRAS.410.1251N} Nardini E., Fabian A.~C., Reis R.~C., 
Walton D.~J., 2011, MNRAS, 410, 1251
\bibitem[\protect\citeauthoryear{Panessa et 
al.}{2006}]{2006A&A...455..173P} Panessa F., Bassani L., Cappi M., Dadina M., Barcons X., Carrera F.~J., Ho L.~C., Iwasawa K., 2006, A\&A, 455, 173
\bibitem[\protect\citeauthoryear{Peterson et 
al.}{2005}]{2005ApJ...632..799P} Peterson B.~M., et al., 2005, ApJ, 632, 
799
\bibitem[\protect\citeauthoryear{Porquet et 
al.}{2004}]{2004A&A...422...85P} Porquet D., Reeves J.~N., O'Brien P., Brinkmann W., 2004, A\&A, 422, 85
\bibitem[\protect\citeauthoryear{Risaliti et 
al.}{2007}]{2007ApJ...659L.111R} Risaliti G., Elvis M., Fabbiano G., Baldi 
A., Zezas A., Salvati M., 2007, ApJ, 659, L111
\bibitem[\protect\citeauthoryear{Risaliti et 
al.}{2009}]{2009MNRAS.393L...1R} Risaliti G., et al., 2009, MNRAS, 393, L1
\bibitem[\protect\citeauthoryear{Risaliti et 
al.}{2011}]{2011MNRAS.410.1027R} Risaliti G., Nardini E., Salvati M., Elvis 
M., Fabbiano G., Maiolino R., Pietrini P., Torricelli-Ciamponi G., 2011, 
MNRAS, 410, 1027
\bibitem[\protect\citeauthoryear{Ross 
\& Fabian}{2005}]{2005MNRAS.358..211R} Ross R.~R., Fabian A.~C., 2005, MNRAS, 358, 211
\bibitem[\protect\citeauthoryear{Shih, Iwasawa, 
\& Fabian}{2003}]{2003MNRAS.341..973S} Shih D.~C., Iwasawa K., Fabian A.~C., 2003, MNRAS, 341, 973
\bibitem[\protect\citeauthoryear{Smith et al.}{2001}]{2001ApJ...556L..91S} 
Smith R.~K., Brickhouse N.~S., Liedahl D.~A., Raymond J.~C., 2001, ApJ, 
556, L91
\bibitem[\protect\citeauthoryear{Vaughan et 
al.}{2005}]{2005MNRAS.356..524V} Vaughan S., Iwasawa K., Fabian A.~C., 
Hayashida K., 2005, MNRAS, 356, 524 
\end{thebibliography}
\end{document}